# Normalization: A Preprocessing Stage

**S. Gopal Krishna Patro[1], Kishore Kumar sahu[2]**

Research Scholar, Department of CSE & IT, VSSUT, Burla, Odisha, India[1]
Assistant Professor, Department of CSE & IT, VSSUT, Burla, Odisha, India[2]

**Abstract**: As we know that the normalization is a pre-processing stage of any type problem statement. Especially normalization takes important role in the field of soft computing, cloud computing etc. for manipulation of data like scale down or scale up the range of data before it becomes used for further stage. There are so many normalization techniques are there namely Min-Max normalization, Z-score normalization and Decimal scaling normalization. So by referring these normalization techniques we are going to propose one new normalization technique namely, Integer Scaling Normalization. And we are going to show our proposed normalization technique using various data sets.

## I. INTRODUCTION

Normalization is scaling technique or a mapping technique or a pre processing stage [1]. Where, we can find new range from an existing one range. It can be helpful for the prediction or forecasting purpose a lot [2]. As we know there are so many ways to predict or forecast but all can vary with each other a lot. So to maintain the large variation of prediction and forecasting the Normalization technique is required to make them closer. But there is some existing normalization techniques as mentioned in my abstract section namely Min-Max, Z-score & Decimal scaling excluding these technique we are presenting new one technique called Integer Scaling technique. This technique comes from the AMZD (Advanced on Min-Max Z-score Decimal scaling) [3-6].

## II. RELATED STUDY

The descriptions of existing normalization methodology are given below:

The technique which provides linear transformation on original range of data is called Min-Mix Normalization [3-6]. The technique which keeps relationship among original data is called Min-Mix Normalization. Min-Max normalization is a simple technique where the technique can specifically fit the data in a pre-defined boundary with a pre-defined boundary.

As per Min-Max normalization technique,

$$A' = \left(\frac{A - \min\ value\ of\ A}{\max\ value\ of\ A - \min\ value\ of\ A}\right) * (D - C) + C$$

Where,
 A' contains Min-Max Normalized data one
 If pre defined boundary is [C, D]
 If A is the range of original data
 & B is the mapped one data then,

The technique which gives the normalized values or range of data from the original unstructured data using the concepts like mean and standard deviation then the

Parameter is called as Z-score Normalization [3-6]. So the unstructured data can be normalized using z-score parameter, as per given formulae:

$$v_i' = \frac{v_i - \bar{E}}{std(E)}$$

Where,
 vi' is Z-score normalized one values.
 vi is value of the row E of ith column
 std (E) = $\sqrt{\frac{1}{(n-1)} \sum_{i=1}^{n}(v_i - \bar{E})^2}$
 $\bar{E} = \frac{1}{n}\sum_{i=1}^{n} v_i$ or mean value

In this technique, suppose we are having five rows namely X Y, Z, U, and V with different variables or columns that are 'n' in each row. So in each row above z-score technique can be applied to calculate the normalized ones. If suppose some row having all the values are identical, so the standard deviation of that row is equal to zero then all values for that row are set to zero. Like that Min-Max normalization the z-score also gives the range of values between 0 and 1.

The technique which provides the range between -1 and 1 is nothing but Decimal Scaling [3-6]. So, as per the decimal scaling technique,

$$v^i = \frac{v}{10^j}$$

Where,
 vi is the scaled values
 v is the range of values
 j is the smallest integer Max(|vi|)<1

But as we all know about these above mentioned techniques well. But the proposed technique one we will discuss in coming section details:

## III. PROPOSED MODEL

As we have studied so many research article, the researchers or scholars who are working in the area of soft computing, data mining etc. and excluding these areas other areas like Image processing, cloud computing etc., of different branches or discipline. If their area of research related to dataset, then must of the dataset are not well structured or dataset are unstructured.

So to make the dataset well structured or make it into the structured one, we proposed one technique, which gives the scaled or transformed or structured or normalized one dataset for our research work within the range 0 and 1.

As like Min-Max, z-score, z-score standard deviation, decimal scaling normalization technique, our proposed normalization technique (AMZD normalization) also gives the range of values between 0 and 1.

Our proposed normalization technique having following features:-
- Individual element scaling or transformation technique.
- Independent of amount of data (large or medium or small data set)
- Independent of size of data (number of digits in each element)
- Scale can be done between 0 and 1.
- Is applicable for integer numbers only.

The proposed normalization technique is given below with explanation;-

$$Y = \frac{(|X|) - (10^{n-1}) * (|A|)}{10^{n-1}}$$

Where,
    X, is the particular data element
    N, is the number of digits in element X
    A, is the first digit of data element X
    Y, is the scaled one value between 0 and 1

This proposed model can be applicable for any length of data element of the type integer only. Except the features we have mentioned above for our proposed normalization technique following are the similarity among our proposed model with the existing techniques namely Min-Max, z-score & decimal scaling is:

- Unstructured to structure one
- Purpose of scaling or formulation
- All works in the numerical data only.

The comparison study through tabulation and graphical representation is described below. Here we like to compare our technique with the existing one Min-Max technique with different data sets.

Below we are comparing our proposed technique with Min-Max normalization technique through table as well as through graph with different data sets like BSE sensex, NNGC and college enrollment data set.

TABLE I
BSE_SENSEX Data Set [7]

| Sl. No. | Original Data | Min-Max Normalization | Integer Scaling Normalization |
|---|---|---|---|
| 1 | 1229 | 0.0976 | 0.229 |
| 2 | 1264 | 0.129 | 0.264 |
| 3 | 1397 | 0.25 | 0.397 |
| 4 | 1455 | 0.303 | 0.455 |
| 5 | 1483 | 0.3284 | 0.483 |
| 6 | 1523 | 0.385 | 0.523 |
| 7 | 1548 | 0.388 | 0.548 |
| 8 | 1594 | 0.429 | 0.594 |
| 9 | 1670 | 0.498 | 0.670 |
| 10 | 1680 | 0.5076 | 0.680 |

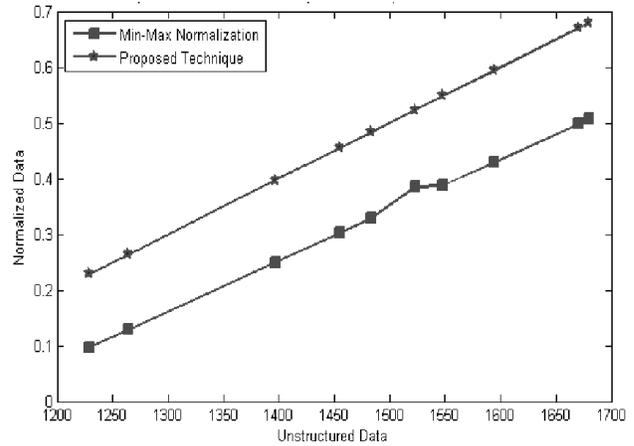

Fig.1 Comparison Graph on Min-Max Vs Proposed Technique for BSE Sensex Dataset

TABLE II
NNGC Data Set [8]

| Sl. No. | Original Data | Min-Max Normalization | Integer Scaling Normalization |
|---|---|---|---|
| 1 | 2677 | 0 | 0.677 |
| 2 | 3083 | 0.062 | 0.083 |
| 3 | 3539 | 0.132 | 0.539 |
| 4 | 4032 | 0.208 | 0.032 |
| 5 | 4452 | 0.273 | 0.452 |
| 6 | 5100 | 0.372 | 0.100 |
| 7 | 5944 | 0.502 | 0.944 |
| 8 | 6913 | 0.651 | 0.913 |
| 9 | 6936 | 0.654 | 0.936 |
| 10 | 9185 | 1 | 0.185 |

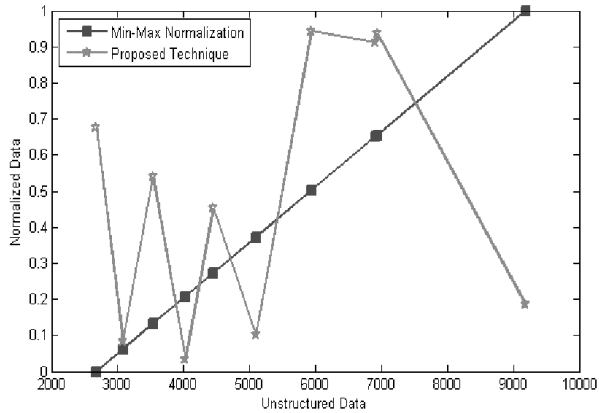

Fig.2 Comparison Graph on Min-Max Vs Proposed Technique for NNGC Dataset

TABLE III
Colleges Enrollment Data Set [9]

| Sl. No. | Original Data | Min-Max Normalization | Integer Scaling Normalization |
|---|---|---|---|
| 1 | 1645 | 0.082 | 0.645 |
| 2 | 2300 | 0.157 | 0.300 |
| 3 | 2472 | 0.176 | 0.472 |
| 4 | 1105 | 0.021 | 0.105 |
| 5 | 7946 | 0.796 | 0.946 |
| 6 | 1657 | 0.084 | 0.657 |
| 7 | 9742 | 1 | 0.742 |
| 8 | 4112 | 0.362 | 0.112 |
| 9 | 917 | 0 | 0.17 |
| 10 | 7219 | 0.714 | 0.219 |

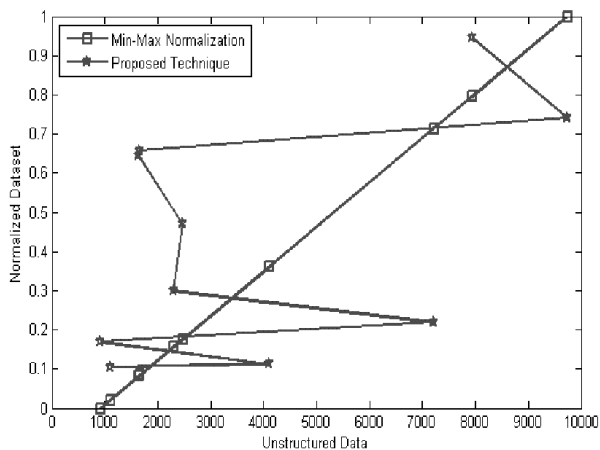

Fig.3 Comparison Graph on Min-Max Vs Proposed Technique for College Enrollment Dataset

In the above tables and graphs, we have described how our proposed normalization techniques works within the range 0 and 1 with respect to the given any range of data.

Following the steps to be followed during normalization:

- Select the range of data of any size.
- Write a code to read that range of data set container file.
- Use proposed technique to scale down range of data into between 0 and 1
- Use the newly generated scaled data into further processing as per our need.
- Then, scale up (if required).
- Finish.

### IV. CONCLUSION

As we have studied that, our normalization technique works well in each and every field of research work like soft computing (we are working), image processing and cloud computing etc. so well,. So we planned it to propose some other types of normalization technique and also use our technique into the fast going research area namely time series financial forecasting as well wherever the data set concept will be arise.

# BIOGRAPHY

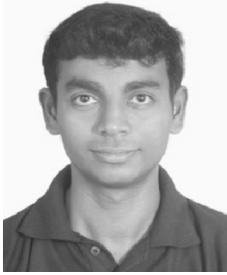

**S.Gopal Krishna Patro** is a M.Tech. research scholar in Department of CSE & IT, VSSUT, Burla with specialization ICT. His area of interest is Financial Forecasting, Machine Learning, and Cloud Computing. He did one international journal in IJCSE. He is having total two year of experience including both industry and teaching

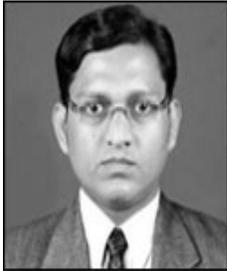

**Mr. Kishore Kumar Sahu** is a Assistant Professor in CSE & IT department, VSSUT, Burla. He is total 10 years of teaching experience at UG & PG level. He is pursuing his Ph. D in Computer Science & Engineering department. His areas of interests are Soft Computing, Artificial Intelligence, Compiler & Theory of Computation.